\UseRawInputEncoding
\documentclass[pdftex,12pt]{article}
\usepackage{amsmath}
\usepackage{amsfonts}
\usepackage{amssymb}
\usepackage{amsthm}
\usepackage{appendix}
\usepackage{graphicx}
\usepackage{enumitem}
\setenumerate{topsep=3pt,itemsep=2pt}
\setitemize{topsep=3pt,itemsep=2pt}

\usepackage{geometry} % Adjusts page margins
\usepackage{color}
\usepackage[colorlinks=true,linkcolor=blue,citecolor=magenta,linktocpage=true]{hyperref}
\usepackage{physics}
\usepackage{bbm}

\newcommand{\f}{\begin{equation}}
\newcommand{\ff}{\end{equation}}

\begin{document}

%%%%%%%%%%%%%%%%%%%%%%%%%%%%%%%%%%%%%%%%%%%%%%%%
\title{ Views, variety and celestial spheres}
%What its like to be \st{an observer} in a quantum universe}
\author{Lee Smolin${}^a$
\\
\\
${}^a$Perimeter Institute for Theoretical Physics,\\
31 Caroline Street North, Waterloo, Ontario N2J 2Y5, Canada\\
and\\
Department of Physics and Astronomy, University of Waterloo\\
}

\date{January 27, 2022}     
\maketitle
  %\vfill

\begin{abstract}

%DRAFT:  NOT FOR DISTRIBUTION

This paper describes a continuation of the program of causal views, in which the world consists of nothing but a vast number of partial views of its past.  Each view is associated to an event, and is a representation of the immediate causal past of that event.  These consists mainly of processes that transfer energy, momentum and other charges to it from its past events.

There is fundamentally no space or spacetime, just a large number of events, which are the causes of events to come.  This is a development of energetic causal set theories, developed with Marina Cortes.   

Momentum and energy are fundamental, and are conserved under their
transformation from present events to future events.    As a result Minkowski spacetime emerges, in a way that preserves causal relations.   The locality of events as constructed in the emergent spacetime is a consequence of the conservation of energy-momentum fundamentally.

In this paper I propose that the views of events can be represented in terms of 
degrees of freedom on punctured two surfaces-each puncture corresponding to an immediate past event.  
%This opens up to the use of mathematical tools from twistor theory and the nascent area of celestial spheres.  We show that 
This makes possible versions of the theory that are relativistically invariant. 

\end{abstract}
\newpage
\tableofcontents

%\newpage

%\tableofcontents

\section{Introduction}

The first  question that physics asks is what the world is ultimately made of.  A measure of the failure of positivism is that, after more than a century of dominance, what we really want to know is still the ontology of the world.  

Newton gave us an ontology of point particles moving in an absolute space, subject to mutual forces at a distance.   While Newton knew better, the sheer usefulness of Newtonian dynamics made it hard to disagree, only a few such as Descarte and Leibniz succeeded
in formulating a viable alternative conceptually - based on the ideas of relationalism.   But even armed with a more consistent philosophy,  none was able to compete with Newton as a paradigm for exact description and prediction.   

But beginning with Faraday and Maxwell, a new ontology did arise that was able to address the questions raised by Newton's action at a distance:  the ontology of fields. And since the turn of the 20th century
fields - both quantum and classical - have dominated our ontological thinking as well as our practices as physicists.

The picture of a universe of fields has inspired many discoveries, but its claims to give a fundamental ontology with a consistent foundation remains aspirational. A few quantum field theories
are known to exist rigorously - and most of these live in $1+1$ dimensions.  The few exceptions are highly constrained by suprsymmetry.   None of
the quantum field theories that make up the standard model are rigorous, in fact our understanding of the physics of quantum electromagntic fields and nuclei
are based on expansions with zero radius of curvature.    Lattice gauge theory appears to be in more hopeful state until we remember that fermions double -
making it impossible to describe those field theories that describe nature.  

There is then a clear need to begin to explore the possibility of a new ontology for physics.  %A very few theorists hav been trying to do this, such as the causal set community, Roger Penrose and David Finkelstein.

This paper reports on a modest attempt to provide an alternative ontology that may be sufficiently
different, and sufficiently rich, conceptually  that it may have a chance to underlie and complete quantum mechanics, quantum field theory and general relativity. 

Called the {\it causal theory of views},   this begins somewhat like the causal set theory in
that we declare that what is real includes events and causal processes, by which they contribute to the catylization of new events\footnote{Related models were proposed in \cite{Cohl,Fotini1}}.  But our events
are both sparser and more ephemeral.  They do not exist in any familiar sense,
 (nothing does, if this new proposal is correct,)   rather we perceive our events to be moments of transformation.  What gets continually processed and transformed, is momentum and energy.  So these have no stable existence-apart from
the very important fact that they are conserved.

An event,  while it lasts, is shaped by those impulses from its causal past-and this gives it a view of its near causal past.   Each view is brief and incomplete.  But the sum of them are the universe.

This is an ontology in which the universe continually recreates itself-as all it is is the
collection of partial views of the causal pasts of the current events.

This paper reports progress in the construction of a theory that is meant to be an ultraviolet completion of general relativity and quantum theory, called the causal theory of
views\cite{CTV1,CTV2}.  
Having shown in previous papers how relativistic point particle dynamics (and also, of course, space)  and non-relativistic quantum
mechanics emerge, in different limits\cite{CTV1, CTV2}, I present here a {\it relativistic theory of causal views.}  By this I mean one whose action and equations of motion are  invariant under lorentz  transformations on momentum spaces\footnote{Of course, because there is no spacetime.  Individual solutions including those that could be candidates to describe our own universe, 
 break those and indeed all symmetries.}.  The key to this is a relativistically invariant expression for the dynamics, based on a invariant measure of the
difference between the views of two events (eqs \ref{halfint}, \ref{Dpart} and
\ref{bmip}, respectively, below).

The main idea is that we live in  a quantum universe  made up of nothing but a vast number of partial views of its  past.  Each view is associated wth an event, and is a representation of the immediate causal past of that event.  

In this theory, space is not fundamental;  nor is spacetime,  although there is a constructive,
Bergsonian notion of time, related to the idea of becoming, or the now, as a consequence of the
continual creation of events.   Energy and momentum and other conserved charges are, however,  present in the
fundamental picture, and each event has so much of each. These are transferred to that event's antecedents in a way that is consistent with their conservation.   These transfers are primarily what an event "can know" about its causal past, and are coded in the views.

 The collection of such views are the be-ables of the theory.    There is a phase of the theory in which non-relativistic quantum theory is derived.   This is the basis for the claim that these theories are non-local 
 hidden\footnote{Despite various rumours to the contrary, there are not a few non-local hidden
 variables theories that reproduce 
 predictions of quantum mechanics\cite{pilot}-\cite{MCU}.}
 variables theories\cite{CTV1,CTV2,real1,real2}.
 
 If space is not present initially, exactly how does it emerge?   We make use of a mechanism
 for the emergence of space first discovered (as far as I know) in the relative
 locality and DSR papers\cite{rl1,rl2,real1,real2} which are also theories formulated in momentum  space.  The idea was then developed in the energetic causal set work.
 
 The key point is simply that conservation of energy-momentum
 implies locality in a spacetime of the same dimension.   We can draw a diagram of every interaction, which shows that processes embedding in a little shard of spacetime.
 The question is how to get the little shards to line up  consistently to define an emergent
 spacetime.  Once put that way we can find the necessary conditions, and construct classes of theories that satisfy them.
 
 At the fundamental level, in the absence of space, how does nature decide which
 interactions are stronger and which are weaker?  
 In \cite{CTV1,CTV2,real1,real2},  a simple answer was proposed:  measures of similarity  of differences of views provides a suitable replacement
 for distant or nearby in space.   So if the most important structure in a field theory is
 the derivative,   in our theory the most basic operation is the comparison of views.
 
 Roughly speaking, the more similar two views are, the more likely they are to interact.
 And the more similar they become, the more repulsive the force\cite{real1,real2,rl1,rl2}.
 
To summarize, we propose that each event in the history of the universe is nothing but the information available there of its  near term causal past.   The more those views differ, the more
diverse the world is.   The universe then is governed by a law which aims to
maximize the diversity of its views\cite{CTV2,JLV}.

This picture is also suggested by the basic twistor duality\cite{twistor}, according to which an event in spacetime is dual to the set of light rays that converge 
there\footnote{However, to incorporate a causal structure we have to break some of the symmetries of twistor space.   We hope to return to this in a future publication.}.

 We also may wonder if there is a relationship between these twistor inspired dualities  and  the celestial sphere\cite{celestial}, recently discussed as a possible structure relevant for quantum gravity.  We show that the new theory proposed here is a kind of inversion of the structure in asymptotically flat  spacetimes, in which we abolish the external boundary, along with its one fictional global outside observer,  and rediscover the inversion of these structures as the microstructure of each event in a quantum universe.   

 The one line summary of the message of this paper is that  {\it the views of celestial
 spheres, turned inside out by a large conformal transformation\footnote{\cite{FM-inside}}, from an event,  are the right degrees of freedom to describe  closed quantum universes.}   Th right dynamics (ie the one that recovers
 non-relativistic quantum mechanics in an appropriate limit), is to extremize the variety or diversity of the views\cite{JLV}.
 
   This is a very old idea, it can be found, for example, in Leibiz's revolutionary  The Monadology\cite{Monadology}.
   
Newton, who was Liebniz's contemporary,  had what would be regarded today as a more conventional view: he imagined that the universe was made of mass points flying around at the command of local equations of motion\cite{Principia}.  
There was one Observer who lived outside the universe and observed it without being observed. 

%We all know how that went.  It worked well for quite a long time, but ran into trouble later on, on issues from giving a  correct definition of inertia that makes Newton's laws non-circular,  to defining consistent notions of point masses and charges; it also troubledthe foundations of quantum mechanics and quantum cosmology.

That archaic picture of a universe that can be seen as a whole by an  all powerful  asymptotic observer, is clearly holding back  progress.

%still colours the imaginations of many theorists who feel they
%can get their footing only in asymptopia\footnote{Why else would you be so outraged
%by the idea of loosing information down the throat of a black hole or across a 
%cosmological horizon.}.  

%I propose here going back to the Monadology\cite{Monadology}  

The universe is not something that exists that we monitor.   To be an observer is, I propose, the same thing as being part of the
universe.  To observe the universe you must be a participant in it.
 
% We are inspired by Wheeler's vision, and we insist with him that in a realist approach to quantum mechanics, there can be no fundamental distinction  between observing and participating\footnote{This has implicationsfor th large scale structure of spacetime, which we hope to return to in a later publication}.   
  
By a view, I mean the information about the causal past of an event, which is coded in degrees of freedom at the event.  But if an event has the puny structure of a "structureless point," this is no more than the values of a few fields, evaluated at that point. There is not much one can do with that.      I find that Leibniz's idea starts to get interesting when  we blow each event up to a two-sphere, so that it has directional information built into it. 

When I use the word ``information" here I do not mean Shannon\cite{Shannon} 
or von Neumann \cite{vonN}  information.   I mean Gregory Batesman information\cite{GB}, defined as "the difference that makes a difference\footnote{Gregory Bateson was an anthropologist and psychiatrist who also invented the concept of the double bind.}."

Here is the basic idea of Bateson:  
We consider making a minimal and local change in the value of some physical observable, $x$  at a time, $t_0$ from $x_1$ to $x_2$.   A long time, $T$
later, we  compare the two evolutions, the one starting  from $x_2$ to that from $x_1$.  

In many cases the late time states are minimally different.   It may be impossible to tell the two apart.  This is typical for systems in thermal equilibrium.   In this case no 
information has been conveyed.   But there are other cases in which the two states are
macroscopically distinct.  Think of a light switch. We say that initial value of $x$ carried information.  The
change from $x_1$ to $x_2$ was a difference that made a difference\footnote{
In statistical mechanics we speak of  damage being done.}.

So what I mean by a quantum universe is a continually becoming  causally related sequence of views of the causal pasts of events, where the view of event, which is equivalent to the name of the event,  is a snap shot of the information arriving from the causal past to the two sphere of directions. 

%But lets be clear that there is only one notion of the past-that based on causation.  So arriving at the $S^2$, {\it from the past} is redundent and means the same as
%just arriving at the two-sphere.

 Furthermore the information coded  on the $S^2$ about its causal past
 could be ``classical", such as the direction  a certain pulse of energy-momentum is arriving from the past,
or a more ``quantum" description, such as a Bell state expressing entanglement of different regions of the past.   W explore both in the third chapter below.

Interpretational issues, concerned with the nature of time,  the structure of a theory of a closed system and quantum foundations, are discussed in \cite{CL1,CL2,CL3}.    As in other cases, the theory does not dictate its interpretation.  In many instances views on foundational issues are not strongly constrained by the formalism, so you can describe these models using the concepts of which ever interpretation you prefer.

We proceed to review the main idea of the $CTV$.
   
   \section{Summary of the causal theory of views\cite{CTV1,CTV2}}.
   
   It is important to emphasize that the Causal Theory of Views is first of all a new proposal for the ontology of the physical world.  
 
 \begin{itemize}

 \item{}The universe is a process of continual becomings and transformations, 
 made up of events and causal processes.
 
  \item{}There is fundamentally no space, no spacetime.   

 \item{}Causal processes carry impulses of energy, momentum and other charges.   This requires taking energy and momentum as primitive, operational concepts, that do not need space to define them. Indeed, they have perfectly good operational definitions in terms of calorimiters,    photomultiplier tubes etc.         We conjecture an inverse theorem which constructs an emergent dimensionality of space for each conserved quantity.
 
 \item{} In all processes involving transfer of energy-momentum it is conserved.
Thus,

\begin{eqnarray}
& {\bf p}_a^{+  J}  &=   \sum_{\mbox{in } M \in Past(J)}  {\bf p}^{+ J}_{M a} 
\nonumber
\\
 = & {\bf p}_{a}^{- J} & =    \sum_{ \mbox{out  } N \in Fut (J)}  {\bf p}^{- J}_{M a} 
 \nonumber
 \\
%&  &=    \sum_{ \mbox{outgoing} N \in Fut (J)}  {\bf p}^{- J}_{M a} 
\end{eqnarray}

 \item{}Events do not exist, they happen.  They are initiated by the combination of two or more causal processes.  Then they do one thing, which is to reshuffle the
 quanta of energy-momntum they receive and send them forth in new 
 causal processes that will initiate a future event.
 
%At this point some readers object that energy and momentum are defined by Noether's theorem, which gives a construction of a conserved quantity for every continuous  symmetry {\it in a pre-existing background spacetime}. %\item{}But, there are  elementary processes, and these carry quantities of energy and momentum.  

 %\item{}The event then transforms into two or more causal processes, each of which receives a share of the event's endowment.
 
 \item{}These transformations and processes conserve energy, momentum and other 
 charges.
 
 %\footnote{although the conservation laws may be non-linear.}.

 %\section{Events and their views}
 
 %Here is an informal statement of our hypotheses.
 
 \item{}There is just one universe, it happens just once.   
 %When we specify a mathematical descriotion of an event, in some cases we are describing the actual universe.  suspended in time.  In other cases we are displaying a collection of possible next states

 \end{itemize}
 
\subsection{A view may be expressed as a quantum state on a celestial sphere}

We define a four dimensional relativistic energy-momentum space,    
${\cal M} $
\f
{\bf p}_a  = (p_i, p_0 = \epsilon)   \in {\cal M} , \ \ \ \ \ \ i=1,2,3
\ff
$\cal M$ is endowed with a lorentzian metric and connection, which may or may not be
related.  

The, simplest characterization of a view is  merely a set of labeled points on a unit punctured $S^2$.  This can be coded into  an unordered list
 of punctures, each of which is
labeled by the angles from which it approached, $(\bar{w},w)$ together with
the energy $\epsilon$.   Each view is, in this version,
\f
{\cal V}_I = \{ ( \bar{w},w,\epsilon  )_I^\alpha \}   .
\ff

If $q_a$ is a null one-form representing a photon's state, it corresponds to a point on
the celestial sphere
\f
q_a = \left (1+ \bar{w} w,    w+ \bar{w} , -\imath (w-\bar{w} ), 1-\bar{w} w    \right )
\ff

Under a lorentz transformation,   $q_a$  transforms as a vector density 
\f
q_a = \frac{1}{| cw+d |^2 }   \Lambda_a^{\ b} q_b   
\ff
$w$ and $\bar{w}$ transform as
\f
w \rightarrow \frac{a w+ b}{c w + d } , \ \ \ \ \  
 \bar{w} \rightarrow \frac{ \bar{a}  \bar{w}+  \bar{b}}{ \bar{c}  \bar{w} +  \bar{d} } 
\ff
with
\f
ad-bc = 1
\ff

%The next step is to throw away the spacetime, the null geodesics, the common point  $\cal CE$, where they coincide, and kEp only the structure which remains as defining a fundamental event.   Each fundamental event is then characterized as a unit $S^2$ on which are positioned a finite set of $n$ labeled points. The latter fall into two sets, 
incoming $\epsilon > 0$ and outgoing $\epsilon < 0$.

We posit that a variable number of impulses $p^\alpha_{E} $ interact in the formation of an event, $E$; and this requires that the energy-momenta carried by these 
impulses
that form the event,  (which are labeled  $p_a^E$  )   
combine to the single injection of energy and momentum, which endows the event.

Thus, the  view of an event $E$ is taken from its causal past
set, which must include at least the elementary processes, or impulsives, whose
combination initiated the event.
\f
{\cal V}_E  = \{  {\bf p}_a \in  \mbox{past}(E)  \}
\ff

This is governed by a map:
\f
  {\cal M} \times {\cal M} \ldots     {\cal M}    \rightarrow  {\cal M}  \ : \ \ \ 
  \phi_1 \otimes \phi_2 \ldots  \rightarrow     \mu 
\ff

%These  maps might be uniform or random, also linear or not.     They may be
%commutative, associative, or not.

%\subsubsection{Probes}

One can also imagine an event making a {\it probe} of the future by sending out
a number of photons, such that the total energy is preserved.     So the
whole action of an event can be understood as a reorganization of the
energy-momenta incoming from the past to a new way of dividing the total incoming energy.
\f
  {\cal M} \otimes {\cal M} \ldots     {\cal M}    \rightarrow  {\cal M}  
  \rightarrow   {\cal M} \times {\cal M}
  \ : \ \ \ 
  \phi_1 \otimes \phi_2 \ldots  \rightarrow     \mu 
\ff

\section{Relativistic measures of differences of views}

There are different ways we can express the view of an event; each leads to a 
different definition of  variety, and hence  potential energy.

They share a common strategy, which is to use the technique of best matching\cite{bm1,bm2} to define differences between a pair of views.      
So next we have a review.

 \subsection{Best matching and local gauge invariance}
 
Our relational philosophy tells us that observables that depend on arbitrary choice of coordinates cannot be physical.
 One way to accomplish this  is through the method of best matching, developed by Barbour and  Bertotti\cite{bm1,bm2}.
 
 We want to compare the past causal views of a pair of
 events, which we will now call $E$ and $F$, without reference to any frame external to or common to them.
 
 We want to treat as identical views that differ from each other by elements of a
 symmetry group $H$.  
 That is, if $\sigma \in H$ does exist such that there are two views $E$ and $F$ such that,
 \f
 F = \sigma \circ E
 \ff
 we consider the two views as identical, as there is nothing in their common
 internal relations that will distinguish them.
 
 We start with a measure of difference, $ d(E,F)$ between two views.   Then we consider
 how this measure of difference is affected by the relative orientations, positions and motions, of the two.

  Thus what we really want is to vary $\sigma$ to find the  group action on $E$
  that minimizes its distinctiveness from $F$.   We call this a best matching
  configuration.
\f
{\cal D}^{rel}(E,F) = 
% \frac{\int d\sigma }{Vol |H|} 
{\cal D}(E,\sigma^* \circ F)d_{min}
\ff 
 where $\sigma$ represents an action of the group $H$
 on the space of possible views.  
  and $\sigma^*$ is the  element of $H$ that minimizes the difference between the two views.
     
 There are  a number of different strategies for doing this comparison.  Each gives
 us a definition of difference on the set of punctures on the $S^2$, which is to say, in the space of views.

\subsection{The view as an unordered set of incoming energy-momentum }

% I will describe the simplest which is to transform the data for the location and energy carried by an impulse back into vectors cited at the origin.  These are given by.

The simplest view of an event $E$ is  an unordered set of $n_{E}$ incoming energy momentum vectors,
\f
v_{E} = \{    q^{E \alpha}_a \ldots  \}
\ff
 Different events receive different numbers of pulses,  $n_{E}$.
 
 We next find a measure of the difference among pairs of these that is
 invariant under Lorentz transformations and permutations.
 
 Let's start by  comparing two light-like energy-momentum vectors.
 We use the Minkowski metric on momentum space:
 \f
 {\cal D}[q_a, p_b ]  =  -\frac{1}{2}  ( q_a - p_a )^2 =    q_a p_b     g^{ab} 
 \label{try1}
 \ff
 This is Lorentz invariant,  non-negative
 when they are both null incoming or  both null outgoing.  It vanishes
 when $p_a = q^a$, so it gives a meaning to how different the two null
 vectors are.
 
 However this comparison makes use of a common reference frame.    
 That is, when we use the form (\ref{try1}), we are  implicitly thinking of comparing two views, made by a single observer.    That is why we apply the same Poincare  transformation to both of them.  
 
Suppose we are in touch with a large number of observers scattered in spacetime.     
They each from time to time take a snapshot of their past view, and send them to us.  But an evil child has gotten into the collection and stripped them all of their labels, so
what is left is an unlabeled and randomly ordered pile of snap shots.
 Our job is to reconstruct the partial order of the views, so as to have the most likely
history of our universe.
 unordered collection of snap shots taken.   
 
 Best matching is a technique for doing just this.  It gives us a Poincare invariant way to give the snap shots  a partial order in terms of differences and similarities of these views.
  % similar two patterns are, taking into account all the possible ways the photographers of one might have been oriented towards the other.
  
To do this, we need the
 freedom to Poincare transform either one, and take the minimum of the resulting differences
 \f
{\cal D} [q_a |  p_b ]_ {b.m.}  =   
((  \sigma \cdot  q_a )  p_b     g^{ab}  |_{\mbox{minimum value over group action  
$\sigma$}}
 \ff
 For example. we can use $\sigma$ to bring the two momenta onto alignment, so the best matched
 value is  $q^0 p^0$ if they are massive, $0$ if they are massless..
 
 Now let us have $n$ incoming energy-momenta in each view.   We label arbitrarily the energy-momentum vectors belong to the same view by indices $\alpha$ and $\beta$. The best matched value of the distinction is
 
 \f
{\cal D}[   \{ q^\alpha_a  \}  | \{  p^\beta_b \}   ]_ {b.m.}  =  \sum_{\pi(\alpha, \beta  \ldots )}
 | ( \sigma \cdot  q^{\sigma \cdot \alpha }_a ) ,  p_b^{\beta \ldots  }  ]     g^{ab} \ldots        
 |_{\mbox{best match}}
 \label{rbn}
 \ff

We evaluate the difference measure over all possible permutations  of one set, and, for each permutation,
we scan the 
group on that side, looking for a set of pairings
and orientations that lead to the greatest matching.  We then define the value of the best matched \cite{bm1,bm2}   to be the minimal difference between the two views, compared over all 
Poincare transformations and all permutations.

In the case that the number of snapshots or incoming impulses in each view are not equal, we consider all ways to distribute the smaller set amongst the larger.

\f
\boxed{
{\cal D}(E,F) = {\cal D} [   \{ q^\alpha_a  \}  | \{  p^\beta_b \}   ]_ {b.m.}  =   \sum_{\pi(\alpha, \beta  \ldots )}
 | ( \sigma \cdot  q^{\sigma \cdot \alpha }_a ) ,  p_b^{\beta \ldots  }  ]     g^{ab} \ldots        
 |_{\mbox{bm2}}}
 \label{Dpart}
\ff

 \subsection{The view as a state in a boundary Chern-Simons theory.}

When we go from the classical theory to the quantum theory the values of
the degrees of freedom are promoted to functionals of possible values.
So, in the "classical" theory we have just described, the incoming  causal processes 
arrives from the past at an event, $E$, from a direction specified by a point $(\bar{w},w)$ on the $S^2$.   In the previous picture this corresponds to a null vector in Minkowski
space, and the  full view from $E$ is the unordered collection of these null vectors.

The best matching  classes of the collections of null vectors give as we ust showed
a language for views suitable for constructing relativistic equivalence classes of views.

If we take the non-relativistic limit that set of vectors reduces to a set of timelike vectors- and these were the basis for demonstrating the emergence of non-relativistic quantum mechanics in \cite{CTV1,CTV2}.  

Next we construct a way to recognize the views that allows us to keep more
topological information. The basic idea is that the set of null directions
$(\bar{w},w, \epsilon)_I$  now become punctures, which we then use to parameterize a
set of Hilbert spaces on the punctured two spheres.    

Note that I will call the first correspondence "classical" and the second "quantum",
but they are both elements of the construction of a classical theory.

The second class of  states assign to each state  a wavefunction  $\psi (z) $ on the 
punctured $S^2$.  From a non-relativistic
perspective, we could expand these functions
in terms of representations of $SO(3)$.   So this would seem to bring us back
to the $SU(2)$ spin networks of $LQG$\cite{LQG-area}.
\f
(\bar{w},w)  \mbox{on the } S^2 )   \rightarrow (j,m) 
\ff
But we can do more.  With the understanding that the conformal two sphere carries
a representation of the $3+1$ Lorentz group, whose action is isomophic to the
$2$ dimensional conformal group, we may expand the wavefunctions in terms of $SL(2,C)$
representation
labelled spin networks.There are 
both compact and non-compact representations   %\cite{noncirp}

But we shouldn't stop here.   To encode the magnitude of the energy-momentum
impulses we need to represent the energy $\epsilon$ in terms of a further
extension, to the Poincare group representation theory.

We want to add the translations so the $\epsilon$'s now indicate how the
states are elements of a  representation of translations.   So we have gone from
\f
(\bar{w},w, \epsilon )  \mbox{on the } S^2    \rightarrow (j,m, \epsilon) 
\ff

which are representation labels for the $3+1$ dimensional Poincare group.   

So we may have a spin foam model constructed from the representation theory of $Poincare (1,3)$, as described in \cite{Poinrep}.

The events  are now intertwinors of $Poincare (1,3)$, with a specified number of
incoming and outgoing zero mass  states.

%The views of the event $E$ then are element of the Hilbert space.
%\f
%{\cal V}_{E} \in {\cal H}_{S^2_{\mbox{punctures}^{Poin (3,1)}
%\ff

We  each have a view, or views, one for each at each moment.  This is precisely information about
our recent causal past, projected onto a sphere in momentum space.  
Our view, is in fact a two
sphere on which is projected photons and other quanta coming from our causal past.  
So it is very natural to formulate a theory of views, as we have sketched it here,
as a theory in which each event is blown up to a sphere.

\subsection{The ladder of dimensions}

Indeed, this idea sits right in the centre of some of the most important ideas about quantum
spacetime geometry to have been studied these last few decades.

I am referring to that very influential concept in quantum gravity and combinatorial topology which is called
the ladder of dimensions.  The basic idea, enunciated by Crane[s\cite{ladder}, 
is  as follows.

The top dimension - in our case $4$  - is a topological quantum field theory on
a four-manifold $\cal M$.  This will be a $BF$  theory for some Lie group (or quantum
group), $G$.

We are interested in the case where ${\cal M}= \Sigma \times R$

There are no non-dynamical fields
on $\cal M$-because the partition function of the $BF$ theory can be shown to be 
independent of the choice of triangulation used to regulate it.   So there are
very few bulk physical observables.

Down one dimension, however, there  are induced physical degrees of freedom
living on the boundary,   
\f
\partial(  \Sigma \times R )=(  \partial \Sigma ) \times R  + \Sigma_+ - \Sigma_-
\ff
These are described by a $G$ Chern-Simons theory on the three 
dimensional boundaries\cite{linking}.  

The two manifold $\Sigma$ may have boundaries as well.  Or we may introduce two
dimensional  boundaries by choosing a surface $\cal B$ which splits $\Sigma$
into two.     The axioms of Atiyah and Segal\cite{Segal-axioms} for topological quantum field theory posit that
for each  embedding of a two dimensional surface $\cal B$ into a compact three manifold, $\Sigma$, that splits that three manifold into
two halves,   $\Sigma_+$ and $\Sigma_-$    there is a Hilbert space,   
$H_{\cal B}$.   For every topological three manifold
$ \Gamma^+$ that has $\cal B$ as its boundary,  ie ${\cal B} = \partial \Sigma^+$, there
is a state $\Psi \in H_{\cal B}$.   Let us assume that $\cal B$ is oriented so it has
two sides, ${\cal B}^+$ and ${\cal B}^-$.    There is a conjugate map, $\dagger$
\f
\dagger : {\cal B}^+ \rightarrow {\cal B}^- 
\ff
We can use $\dagger$ to construct an inner product, given 
a state $ |\psi > \in H_{\cal B}^+$, and a dual state,  $    < \phi | \in H_{\cal B}^-$,
the inner product is naturally defined 
\f
< \psi | \phi > \in C
\ff
This is then an invariant of the topology of $\Sigma$ because it is invariant under the choice of the splitting surface, $\cal B$.

Let us call this structure  the  ladder theory. 
The ladder theory is not quantum gravity-nor is it a model of quantum gravity in four
spacetime dimensions. All of its degrees of freedom and physical observables
live on the three dimensional boundary.  There is what has been called a {\it boundary 
observables algebra}\cite{linking}.

Very remarkably quantum gravity in three dimensions
is a TQFT-as was first discovered by Ponzonno and Regge\cite{PR}.

Even more remarkably, there are several ways that the ladder theory may be disordered or constrained 
to yield a quantization of general relativity in four spacetime dimensions with
three dimensional boundaries.    

\subsection{Quantum gravity and spin networks}

One way to do this is to disorder the
partition function by embedding spin  networks for $G$ into the 3 dimensional 
bulk\cite{SN-Roger,SN-LQG},
$\Sigma$\footnote{A $G$ spin network is an abstract graph, whose edges are labeled
by irreducible representations of $G$ and whose vertices are labeled by
intertwiners of the incident representations.  When embedded in a manifold as just
described it is also called a spin network.}.  These end with punctures on the $2$ dimensional boundary.  These inherit the labels on the  spin network graphs, which are $G$ representations.   That is to say we extend the previous correspondences.
Now we assign a Hilbert space, $H_{{\cal B}_{p^\alpha, j_\alpha }}$ to  every punctured two surface, ${\cal B}_{p^\alpha, j_\alpha }$.  And a state in that Hilbert space is assigned to every embedding of a spin network in $\Sigma^+$ that ends on the punctures,
matching irreducible representations, $j$-  up to topological deformations of the embeddings that leave the
punctures fixed.

Given the correspondence arising from $QG$ between representations 
and intertwinors of $SU(2)$ and area and  volume these states have natural
interpretations as triangulated three geometries.

When $G$ is chosen to be a local symmetry or gauge group for general
relativity, the resulting partition function can be shown to be a quantization of general relativity\cite{MM,artemMM,linking,Plebanski}.
These are  called spin-foam models.  One of the things that is wonderful about them is that one can continue to use the boundary observables.   In some cases they inherit new meanings from the map between the TQFT and gravity.   For example the non-Abelian electric fluxes now carry geometrical observables such as areas and volumes.

In some cases the partition functions have ambiguities; this can happen because there are physical observables that depend on chirality, which is not represented by the 
one-dimensional structures of the spin networks.  These 
 require modification of the spin networks.   The edges of the spin networks may be blown up into tubes.  There are new degrees of freedom which are winding numbers
of the tubes.  

The end points of those lines, on the spatial boundaries, which were points, are now blown up into circles.       To represent these we must 
extend $G$ to give it structure that can transform non-trivially under diffeomorphisms of these circles as well as under parity
transformations.   

When $G$ is compact we can extend it to a quantum group at root of unity.
Or we extend the group to a loop algebra (or Kac-Moody algebra) the group of mappings of a circle into a Lie group.     The Virasoro group, which is the group of mappings of
circles onto circles, centrally extended,
acts at those circles because it is a subgroup of the extension of $G$, giving us new physical degrees of freedom, which can be represented by a $CFT$ such as 
the $WZW$ theory.   

This structure fits nicely into the triangulations of four dimensional manifolds
that we use in quantum gravity.  The dual of the triangulation is a graph in a three dimensional bulk. The tetrahedra form the
bulk and dual to each tetrahedron is a three dimensional spin network.  Each 
tetrahedron is dual to  an intertwiner.  Its four triangles  each are dual to a shared
boundary of two tetrahehedra, which in turn is dual to an edge connecting the node in each
tetrahedra.  Every four-simplex is bounded by $5$ tetrahedra.   That four simplex then represents an event in which $n$ tetrahedra are replaced by the  $5-n$ tetrahedra.

In this way we get a quantum theory of gravity based on a lorentz group
gauge symmetry, $G= SL(2,C)$ or $SO(3,1)$.  This is close, but it is not exactly what we want.    The models based on dynamical causal structure need to label the punctures by relativistic four momenta, $p_a^I$.   So do the celestial spheres.

There are at least two ways to do this

\begin{itemize}

\item{} Extend the gauge group to the Poincare group.

This requires a good dose of the infinite dimensional unitary representations of
the Lorentz and  Poincare groups, so we postpone further discussion of it to a later publication.

\item{} Extend the group to the deSitter or Anti de Sitter group.
\f
G = SL(2,C)   \rightarrow SU(2,2)
\ff

Construct the theory for the deSitter group.  Then take $\Lambda $ to
zero, if needed.  

\end{itemize}

\subsection{The Chern-Simons boundary action}

The second road is technically simpler, so we start with that:
  We work in a spinorial version of the MacDowell-Mansouri formulation of general 
relativity\cite{MM}.  Everything we do we work with Lorentz spacetime.   This is a small modification of a  $B \wedge F$ 
theory\cite{ArtemMM} for the double cover of the
Desitter (or Ads) group:   $SU(2,2)$
%\footnote{Gravity theories which depend on the holonomy of an $S(2,3)$ connection in this way have  been studied previously in{artem3}.}
   
   The novelty of this formalism is that we add a twist to the idea that general relativity is a constrained topological field theory, by making that theory a gauge theory of $SU(2,2)$,  where the constraints that introduce local degrees of freedom break the gauge group  down to $SL(2,C)$.   

As a result, 
the frame field one form $e^{AA'} $ is expressed as components of the $SU(2,2)$ connection,
in $\frac{SU(2,2)}{SL(2,C)}$   
that exists because $SU(2 ,2)$  is broken spontaneously to $SL(2,C)$.
i.e. the metric geometry is a Higgs field- an order parameter that marks the spontaneous breaking of $SU(2,2)$ to $SL(2,C)$.

%order paramerer and symmetry breaking.
 
 The implications of this extension can be worked out in any first order version or extension of general relativity.  
 For the convenience of the reader who may not be familiar 
with the exotic Plebanski formalism\cite{Plebanski}, and is more likely to have
met the Palatini variables\cite{Palatini}.
 We work here with the latter variables.
 
  The theory is based on an $SU(2,2)$ connection, ${\cal A}^{IJ} $,  which decomposes as,
\begin{eqnarray}
{\cal A}^{AB} & = &  {\cal A}^{(AB)}  = A^{AB}    \ \ \ \ \ \ \  \mbox{chiral $SU(2)_L$  connection}
\nonumber
\\
{\cal A}^{A'B' }  & = & {\cal A}^{(A'B')}  = A^{A'B'}  \ \ \ \ \ \mbox{chiral $SU(2)_R$  connection}
\\
{\cal A}^{A A'}    & = &   \sqrt{\Lambda }   \ \ \ \ \ \ \ e^{A A'}  \ \ \ \ \ \ \ \ \mbox{the frame field
one-forms}
\end{eqnarray}

We see that this approach requires a non-vanishing cosmological constant,
 $\Lambda  $\footnote{  The $I, J \ldots= (A A') = (010' 1')$ are four component Dirac spinor indices, and we are taking advantage
of the two to one map between $SO(5)$ and $SU(2,2)$.}.   

 The corresponding components of the curvatures are
\begin{eqnarray}
{\cal F}^{AB} & = &   dA + A^2   + \Lambda e^{AA'} \wedge e^B_{A'}
\\
{\cal F}^{A'B' }  & = & d A^{A'B'}   + A^{A'B' } \wedge A   
\Lambda e^{AA'}   \wedge e^{B'}_{A}
\\
{\cal F}^{A A'}    & = &   \sqrt{\Lambda}  
(   d e^{A A'}    +   A^{A}_{B} \wedge e^{BA'}   +   A^{A'}_{B'} \wedge e^{A B'}         ) 
= {\cal D} \wedge e^{AA'}
= {\cal T}^{AA'}
\end{eqnarray}

where ${\cal T}^{AA '} $ is the torsion tensor.  

The action of the $SU(2,2)$  $BF$ topological theory is

\begin{eqnarray}
 S &=& - \imath \int_{\Sigma \times R} \left (
 \frac{1}{g^2} B^{IJ} \wedge {\cal F}_{IJ} - \frac{1}{e^2} {\cal F}^{IJ} \wedge {\cal F}_{IJ} 
 %-- \frac{1}{e^2} - \frac{1}{f^2}  {\cal F}^{IJ} \wedge {\cal F}_{IJ} 
 %one more
 \right )
 + \int_{\partial \Sigma\times R}  \frac{k}{4\pi}  Y_{CS} (SU(2,2))
  \nonumber
  \\
& = & - \imath \int_{\Sigma \times R} 
\frac{1}{g^2} \left [
  B^{AB} \wedge {F}_{AB} -  B^{A'B'} \wedge {F}_{A'B'}   
  -B^{AB'} \wedge {F}_{AB'} 
  - {1}{n}^2 
  +  %\frac{1}{g^2} 
  B^{AA'} \wedge {\cal D} e_{AA'} \right  ]
 % \nonumber
  \\
 & -&  \frac{1}{e^2} (  F^{AB} \wedge F_{AB} +F^{A'B'} \wedge F_{A'B'} +
 {\cal D} e^{AA'}   \wedge  {\cal D}  e_{AA'}     )
 % \nonumber
 % \\
  %&&   
  +\Lambda   e^4  [\frac{4}{g^2}  + \frac{4}{e^2}   ]
  \nonumber
  \\
  & +&  \int_{\partial \Sigma\times R} \frac{k}{4\pi}  (Y_{CS} (SU(2)_L)
  -  \frac{k}{4\pi} Y_{CS} (SU(2)_R)  +
    \frac{k}{4\pi} e^{AA'} \wedge {\cal D} e_{AA'} )
  \nonumber
\end{eqnarray}

The boundary term is
\f
S_{\cal B}^{BCS} = \frac{k}{4\pi} \int_{S^2 \times R}
\left (
Y_{CS} ( SU(2)_L ) -  Y_{CS} (SU (2 )_R)   + e^{AA'} \wedge  {\cal D} e_{AA'}
\right )
\ff

One breaks the symmetry by adding Lagrange multipliers which enforce
\f
B^{AB} = e^{AA'} \wedge e^B_{A'}
\ff

After we break the symmetry by implementing these constraints,   the action is then the Palatini action plus topological  and boundary terms.
\begin{eqnarray}
 S^{Palatini} &=& -\imath \int_{\Sigma \times R} 
( \frac{1}{g^2} + \frac{4}{e^2}) \{   e^{AA'} \wedge e^B_{A'}  \wedge { F}_{AB} 
  -   e^{A' A} \wedge   e^{B'}_A  \wedge F_{A'B'} \}  
    - \Lambda e^4 (\frac{2}{g^2}+  \frac{4}{e^2} )  +  \frac{1}{g^2}B^{AA'} \wedge {\cal D} e_{AA'}   ]
   \nonumber
   \\
&&   + \frac{e^2}{2} ( F^{AB} \wedge F_{AB}   +  F^{A'B'} \wedge F_{A'B'}
   +   (  {\cal D} e^{AA'} \wedge {\cal D} e_{AA'} )   
   \nonumber
  \\
    && +  \int_{\partial \Sigma \times R}   (      \frac{k}{4\pi } Y_{CS} (SU(2)_L)
  - \frac{k}{4\pi }Y_{CS} (SU(2)_R)  +   \frac{k}{4\pi }  e^{AA'} \wedge {\cal D} e_{AA'} )
\end{eqnarray}

Let us look  at the first variation and make sure the action is functionally differentiable.
\begin{eqnarray}
\delta S & = &  \int_{\cal M}  [       (EOM)_{AB} ] \delta A^{AB}    
+  (EOM)_{A'B'}  \delta A^{A'B'}  
+(EOM)_{AA'}   \delta e^{AA'}
\nonumber
\\
&+&   \int_{\partial {\cal M}}  ( [ ( \frac{1}{g^2} \Sigma_{AB}   -
(  \frac{1}{e^2}  - \frac{k}{2 \pi})  F_{AB}   ]\wedge \delta A^{AB}  +
( \frac{1}{g^2} \Sigma_{A'B'}   -
(  \frac{1}{e^2}  - \frac{k}{2 \pi})  F_{A'B'}   ]\wedge \delta A^{A'B'}
\nonumber
\\
&+&[ \frac{k}{2\pi}  -\frac{1}{e^2}     ] \ \delta e_{AA'} \wedge {\cal D} \wedge e_{AA'}  
\end{eqnarray}

For the action to be functionally differentiable, the boundary term of
the variation must vanish.   

As was first shown in \cite{linking,me-holo,meyl-holo} there are a number
of ways to accomplish this.  There are subtleties in each signature.

We discuss here only the  basics of the Lorentzian signature\cite{me-holo}.

There is a set of boundary conditions that leave the edge modes for
$\epsilon^{AA'}$, $a^{AB}$ and $a^{A'B'}$,
 (the pull back of the forms into the timeike or null boundaries)
 all free,
 \begin{eqnarray}
 \frac{1}{g^2} \Sigma_{AB}   &= &
 \frac{1}{e^2}  - \frac{k}{2 \pi})  F_{AB}   
\\
 \frac{1}{g^2} \Sigma_{A'B'}   & = &
( \frac{1}{e^2}  - \frac{k}{2 \pi})  F_{A'B'}   ) 
\nonumber
\\
0 &= &
( \frac{1}{e^2}  - \frac{k}{2 \pi})  {\cal D} \wedge e_{AA'}  
 \end{eqnarray}

This affirms that we must have a boundary term which includes the energy and momentum,
i.e. one based on the Poincare or deSitter groups.

 The translations are generated by energy and momentum, thus we can label
 the punctures by energy and momentum, and complete the picture of a causal theory
 of views based on celestial spheres.

We then have on each event's $S^2$ a set of $ n= n_{in} + n_{out} $ punctures,  of which $n_{in}$ are
incoming  and $n_{out}$ outgoing.  

The quantum field theory on each $S^2_{p_\alpha, j_\alpha}$'s is based on a Hilbert space
${\cal H} _{n_{in},n_{out} (\bar{\sigma} \sigma )_I \ldots } $.   We work with a
left-right symmetric spin network basis\cite{me-holo}, where the
edges are labeled by representations of $SU_L (2), SU_R (2)$.

The irreducible reps are labeled by $(j_L, j_R, I_L , I_R )$.    
The reality conditions impose the physical representations are
balanced $j_L = j_R  $.

\subsection{Some remarks}

The next step is to define a difference operator $\hat{\cal D}(E,F) $
on 
\f
\hat{\cal D} :
{\cal H}_E \otimes {\cal H}_F  \rightarrow    R_+     .
\ff
We can use the inner product on Chern-Simons theory to measure similarity.  A state of $G$ Chern-Simons theory on the $n_A$ punctured $S^2$, notated,
\f 
| A, n_A, (\bar{y}, y_i , r_i ) >
\ff
lives in a Hilbert space ${\cal H}_{n_A, (\bar{y}, y_i , r_i )}$.    

This unphysical Hilbert space is dependent on the positions on the $S^2$'s of the
$n_A$ punctures.   

We note that the punctures play a role analogous to the $n$ particle trajectories
in the particle construction.    As in that case we ask for the best matching to
construct the inner product.   
The Hilbert spaces decompose according to the positive number of punctures $n_{A}$
\f
{\cal H}_{   {\cal S}_{p_\alpha, j_\alpha} }= 
\int_{S^2_{p_\alpha, j_\alpha}} (\bar{dy}, dy ) {\cal H}_{{\cal S} - punctures} (\bar{dy}, dy )
\ff
We then define the inner product by
\f 
<  B, n_B r_j   | A, n_A,  , r_i ) >  =   <B|A> \delta_{n_A n_B }  \delta_{r_j r_k }
\ff
the positions of the punctures don't matter up to diffeomorphisms of $S^2 $- punctures.
This is reflected in a braid group symmetry   ${\cal Br}(n_A )$ acting on the Hilbert
space ${\cal H}_{n_A}$.

If the group $G$ were compact, these Hilbert spaces would be of finite
dimension.    As it is for the lorentzian theory we would probably be constrained to
work within the countably infinite unitary representations\cite{JM}.

The Hilbert spaces decompose according to the positive number of punctures $n_{A}$

\f
<   A ,n_A, ( y_i , r_i )| \phi \circ\ \phi \circ | B, n_B ,  ( y_i , r_i )   >  =
  <B|A> \delta_{n_A n_B }  \delta_{r_j r_k }^{bm.}
\ff

\f 
<  B, n_B r_j   | A, n_A,  , r_i ) >  =   <B|A> \delta_{n_A n_B }  \delta_{r_j r_k }
\ff
We can then define the difference in two views, view $A$ on $n^A$ punctures, which are in the past of $E$,  and view $B$ on $n_B$ punctures.  For a quantum system it is
natural to take the inverse of the inner product to be the difference.  As in the classical
description,  we compare first the punctures, and then the fields.
\f
{\cal D}[E, F] = \frac{1}{<\psi_E \ \psi_F >}|^2
\label{bmip}
\ff

 \section{The dynamics}

If the context is novel, the dynamics will be as much as possible conventional.
We begin by defining potential  energy, then an Hamiltonian.  

\subsection{The variety as potential energy}

To make the potential energy,  we sum 
the views- difference   ${\cal D}[E,F]$ 
over all  pairs of causally unrelated events,$<>$.   Because it is based on a best
matching procedure,  which takes the best value over all pairs in the orbit
of the Lorentz group action ${\cal D}[E,F]$ the result is lorentz invariant.   That a pair is not causally related is also Lorentz invariant.  Thus, up to divergent terms, and effects of non-invariant cut offs, the potential energy is lorentz invariant.

\f
U =\sum_{I <> J } {\cal D}(I,J)_{best matched}
\ff

The last step is to restrict to events $I$ and $J$ in the past of a third
{``observer event"}, $K$.
\f
{\cal D}(I,J;K)_{best matched}
\ff

We then sum over all triples 
\f
{\cal V}_{<>} = \sum_K  \sum_{I <> J |in \mbox{Past}(K)}  {\cal D}(I,J;K)_{best matched}
\ff

We propose to call this the {\it acausal variety.}   We will propose it as
a measure of potential energy.   
\f
g {\cal U}  =   {\cal V}_{><} = \sum_K  \sum_{I >< J | \in \mbox{Past}(K)}  
{\cal D}(I,J;K)_{best matched}
\ff
$g$ is a coupling constant.  

We notice that the best matched value of a comparison is lorentz invariant, as it considers the differences scanning over the whole group.   Additionally the 
criterion we use to  pick out which pairs of events to be compared are not
altered under the operations defined in. 

One might want to call the potential energy non-local.
I prefer to call it a-local because it makes no reference to space.    The theory does have an emergent spatial geometry and relative to that
the potential that derives from the variety remains non-local, as is discussed in \cite{CTV1,CTV2}.

The Hamiltonian is then
\f
\boxed{
{\cal H}_0 =  g {\cal U} }
\ff

In the non-relativistic limit, defined as kinetic energy dominated by some 
mass, $m$, this exactly reproduces Bohm's potential\cite{Bohm}.   

Note that in the non-relativistic limit a kinetic energy term will appear
proportional to\cite{CTV1,CTV2}.   In the full relativistic theory the kinetic
energy terms are hidden in the constraints.
\f
K.E. = \frac{(p_i )(p_j )g^{ij}}{M}
\ff

\subsection{The dynamics of difference: the half- integral}

We are ready now to define the dynamics of the theory.
%\footnote{We will assume that the momenta and energies all live in a commutative subalgebra, so we can integrate or sum over them.}.
%\f
%{\cal Z}[\tau ,  \tau , {\cal J}]  = \sum_n} \int    \ff

We propose that the dynamics is defined by the following {\it half integral}.

The degrees of freedom that we integrate over are the causal structure, given by the
partial order structure on the $S^2$'s and the energy-momentum transferred
in each causal relation.     

Imposing the constraints as $\delta$ functionals in the
measure of the half integral.
\f    
\boxed{
{\cal Z}(J)= \sum_{processes}    [\prod_{I  |> J}  dp_{a I}^{J}  
\delta ( {\cal C }_I^J ) ) \delta  ( {\cal Q}^I_J   )  ]
 \prod_I   \delta  ( {\cal P }_a^I )
 e^{\imath    {\cal H}_0 (p)  }}
 \label{halfint}
\ff

The constraints are imposed as constraints in the measure
of the half-integral.

This is the complete definition of  the theory.
There is no reference to space or spacetime at the level of the fundamental
definition of the theory.
As a consequence there is no $\hbar$.
There are no non-trivial commutation relations
and
no uncertainty principle.

%\subsection{Exponentiating the constraints}

\subsection{ Semiclassical limit:  the emergence of Minkowski spacetime}

Now we introduce Lagrange multipliers to exponentiate each of the constraints.
There is a conservation law ${\cal P}_a^I$ for each event, so we write,
\f
 \delta  ( {\cal P }_a^I )  = \int  dz^a_I e^{\imath z_I^a  {\cal P }_\alpha^I  }
\ff

The next step is finding the equations of motion.   We will se that the semiclassical
theory is sufficient
to understanding  how a classical spacetime emerges
from this theory\footnote{we assume the  simplest case where the momentum is conserved linearly},
\f
{\cal P}^I_a  = \sum_{J < I} p_{a I}^J - \sum_{K > I} p_{a K}^I
\ff

Take the first variation by
$p_{aJ}^K$ of the action and set it to zero.    
\f
0 =\frac{\delta S^{eff}}{\delta p_{a J}^K}
=  ( z^a_J - z^a_K ) +N p_b^I   g^{ab}      +g \frac{\delta {\cal V}}{\delta p_{a J}^K}
\ff
where $g_{ab}$ is the metric on momentm space.

%\frac{ {\cal P}}{\delta p_J^K}_a^I  
%+ N \left [ \frac{ {\delta \cal P}}{\delta p_J^K}     + \frac{{\cal P}}{\delta  p_J^K}     \right ]

At first,  ignore the potential term
\f
g \frac{\delta  {\cal V} }{\delta p_{a^J}K}
\ff

It then follows that in the limit $g \rightarrow 0$, 
\f
 ( z^a_J - z^a_K ) ( z^b_J - z^b_K ) g_{ab} =0
\ff

We see that the Lagrange multipliers $z^a$ have become coordinates on an
emergent Minkowski spacetime.
We see also that $g_{ab}$ is a conformal metric on spacetime and the
intervals  $z^a_J - z^a_K$   are null.  So we see that a conformal metric emerges on
spacetime, which is just the inverse of the metric on momentum space.

 \section{Conclusions}

Any proposal for a new physical ontology, faces  huge challenges, even greater that
those that confront attempts to understand quantum gravity or foundations within
a more standard ontology.    

The causal theory of views was proposed in \cite{CTV1,CTV2}, using elements from previous models of spacetime; causal set theory, energetic causal sets,  and is influenced by relative locality\cite{rl1,rl2}, twistor theory\cite{SN-Roger}, the study of amplitudes
and loop quantum gravity\cite{LQG}.   How is it doing?

\begin{enumerate}

\item{} The framework of the ontology is clear but there are several subtle issues that are still being studied, in \cite{CL1, CL2}.

%e posit that he world is a process of events which give rise to new events through causal processes.  

\item{}The be-ables are views of an event-which is what can influence an event
from  that event's causal past.   We have several versions of this; these models differ by the mathematical framework used.

\item{}Fundamentally there is no space or spacetime.  Energy and momentum
are fundamental; relativistic spacetime emerges as a consequence of the conservation of energy and momentum.   We have seen how this happens in several contexts.

\item{}The dynamics is based on comparisons of how diverse are the
views of pairs of events.   The potential energy is related to the variety of
the universe, which is the sum over all pairs of their differences.

\item{} We understand how to derive non relativistic quantum many body theory,
due to the similarity of the variety and the Bohmian quantum potential.

\item{} We have a start on formulating relativistic theories of views.
We  see how special relativity can emerge readily, making use of the connection
between energy-momentum conservation and locality in spacetime.   But how will
general relativity appear?   One answer is to introduce parallel transporters into the causal processes, so the energy momentum that arrives need not be that which was sent.

\item{} 
The statistical physics of the theory described here is going to be very interesting.
The statistics of the possible and actual states or transitions are very different in our context.   Preliminary numerical studies by M Cortes of a simple model show new phases and new phase transitions\cite{ECS1}-\cite{ECS4}.      The equilibrium ensemble is not reached during the lifetime of the universe.
As a result, our universe is, we conjecture, very far from ergodic.   

%You may think you are approaching equilibrium, but you are getting further and further from an actual static distribution as the rate of invention of novel states with novel properties is growing even faster\cite{biocosmos2}.

\item{} There is an expectation that underlies or motivates the study of asymptotic structures. Spacetime may be really weird when we probe it at very short distances, but if we could travel in an opposite direction,  further and further away from us, at longer and longer wavelengths, it
is ultimately more of the same, only more so.    Our view is quite the opposite:  far far
away from us is going to be more and more quantum and, indeed, beyond quantum.   There is no reason to expect a classical asymptopia,  when 
$IR/UV$ symmetry governs, going to the very very large gets to the same "place" going very very very small.

Which is to say that if we haven't yet seen indications of $IR/UV$ symmetry we have
not yet begun to study real cosmology.

\item{} The Causal Theory of Views is also part of another research program, which is based on the idea that completions of background dependent theories eliminate their background dependence by a process which replaces dualities with trialities.

This is based on an observation that in many cases of dualities in physics, the duality transformations are based on a fulcrum of background structure.

For example, $UV/IR$ dualities or in general weak/strong dualities leave fixed a scale
which defines which is which.   Or the Born dualities are based on a fixed definition of
time.

In these cases a deeper theory was found by
eliminating
the dual pairs' dependence on non-dynamical background elements by elevating the duality to 
a triality  in which each of the three elements is defined by the interaction of the other two\cite{triality-3}.

This motivated my studies of cubic matrix models, which furnish many examples
of such duality to triality moves.    Indeed, string theory furnishes some beautiful
examples of the passage from a background dependent to a background independent
formulation which is cubic.  (Think of the construction of actions for strings of the
form of $S = Tr \Phi^3$.)

There are also many examples in the mathematical general relativity literature of
the special role played by cubic actions, such as the Plebanski action.  These cubic
formulations are by far the simplest - since the actions are cubic the equations of
motion are all quadratic equations.  Many new unexpected results were made
possible by adopting one of these cubic formulations.

\item{} This proposal raises a number of fundamental issues concerning quantum
foundations and the nature of time, which are discussed elsewhere\cite{CL1,CL2,CL3}.

\end{enumerate}

Much remains to do.

\section*{ACKNOWLEDGEMENTS}   I would like to thank Luca Ciambelli, Anna Knorr 
and Sabrina Pasterski for comments on early drafts of this paper; also  Stephon Alexander,  Giovanni Amelino-Camelia, Julian Barbour, Marina Cortes, Louis Crane, Laurent Freidel, Jaron Lanier,  Joao Magueijo, Roberto Mangabeira Unger,  Fotini Markopoulou, Carlo Rovelli, and  Clelia Verde for collaborations on related projects.  I am grateful to David Finkelstein,  Lucien Hardy,   Roger Penrose, Robert Spekkens,  and Antony Valentin for many critical conversations which helped shape this work.

This research was supported in part by Perimeter Institute for Theoretical Physics. Re- search at Perimeter Institute is supported by the Government of Canada through Industry Canada and by the Province of Ontario through the Ministry of Research and Innovation. This research was also partly supported by grants from NSERC, FQXi and the John Templeton Foundation.

\end{document}